\pdfoutput=1
\documentclass[prd,aps,showpacs,preprintnumbers,amsmath,amssymb,nofootinbib]{revtex4}
\usepackage[pdftex]{graphicx}
\usepackage{amsmath}
\usepackage{amssymb}
\usepackage{verbatim} 

\usepackage{graphicx}
\usepackage{dcolumn}
\usepackage{bm}
\def\be{\begin{equation}}
\def\ee{\end{equation}}
\def\bea{\begin{eqnarray}}
\def\eea{\end{eqnarray}}

\begin{document}
\title{Angular 21~cm Power Spectrum of a Scaling Distribution of
Cosmic String Wakes}

\author{Oscar F. Hern\'andez $^{2,1}$ \footnote{oscarh@physics.mcgill.ca},
Yi Wang $^{1}$ \footnote{wangyi@physics.mcgill.ca}, 
Robert Brandenberger$^{1}$ \footnote{rhb@physics.mcgill.ca} and 
Jos\'e Fong $^{3,1}$ \footnote{jose.fong@ens-lyon.fr}}

\affiliation{1) Department of Physics, McGill University,
Montr\'eal, QC, H3A 2T8, Canada}
\affiliation{2)  Marianopolis College, 4873 Westmount Ave., 
Westmount, QC, H3Y 1X9, Canada}
\affiliation{3) Ecole Normale Sup\'erieure, Lyon, France}


\begin{abstract}
Cosmic string wakes lead to a large signal in 21~cm redshift maps at
redshifts larger than that corresponding to reionization. Here, we
compute the angular power spectrum of 21~cm radiation as predicted by a scaling distribution of cosmic strings whose wakes have undergone shock heating. 
\end{abstract}

\maketitle

\section{Introduction}
\label{sec:introduction}
A new observational window to probe the structure of the universe
is opening up: 21~cm redshift surveys. Such surveys offer the
prospect of mapping out the distribution of neutral hydrogen in
the universe, and thus will be able to probe the distribution of matter in the
``dark ages'', before star formation sets in (see \cite{Furlanetto}
for an extensive review). Regions in space which have an overdensity
of baryons will lead to excess emission or absorption signals in
21~cm redshift maps.

In a previous paper \cite{us} we considered the signal of a single
cosmic string in a 21~cm redshift map. Cosmic strings moving through
space produce nonlinear matter overdensities in their wakes - even at
redshifts larger than that of reionization. We pointed out that cosmic
strings which are relevant to cosmological structure formation will
leave a large imprint in 21~cm redshift surveys, in particular at
redshifts larger than that corresponding to reionization, a redshift
range where the ``noise'' from reionization processes is not
present. As will be briefly reviewed below, we found that a single
cosmic string gives a characteristic signal in 21~cm redshift maps: a
wedge which is thin in redshift direction and extended in both angular
directions (see Figure 1). The planar extent of the wedge is set by
the length of the string at the time it is formed, and the wedge
thickness depends on the string tension.  The angular extent of the
signal from a string wake produced at the time of
recombination is of the order of $1^{\circ}$, and the mean thickness
$\delta z$ of the wedge in redshift direction (which depends on the
tension of the string ) is about $\delta z / z \sim 10^{-4}$ if the
string tension $\mu$ is given by $G \mu = 10^{-7}$ ($G$ being Newton's
gravitational constant) and the string is moving with a velocity close
to the speed of light. The brightness temperature of the string signal
depends both on the redshift of string formation and that of 21~cm emission, but only
very mildly on the string tension. 
The 21~cm brightness temperature of a wake at a redshift $z_e+1=20$ can be of 
the order of $-15~\rm{mK}$ for strings formed during matter radiation equality with 
$G \mu =  10^{-7}$. Based on this large signal we
expect that it will be much easier to see the signals of strings - if
they are present - in 21~cm maps than in other cosmological windows of
observation.

There has recently been renewed
interest in the cosmological effects of cosmic strings (see e.g. \cite{VS,HK,RHBrev,Wu:1998mr,Durrer:2001cg}
for reviews on cosmic strings and their effects in cosmology). It has been
realized that many particle physics models beyond the Standard Model of
particle physics give rise to cosmic strings. Such strings arise, for example,
at the end of inflation in many supergravity models \cite{Rachel}. Similarly,
cosmic strings are a generic remnant of many models of brane inflation
\cite{CSbrane}.

Furthermore, if strings exist, there will not be just one.
In models which lead to cosmic strings, a network of
strings will inevitably form in a phase transition in the early universe.
Causality arguments \cite{Kibble} then imply that these strings will persist
at all times up to the present time. Both analytical arguments 
\cite{VS,HK,RHBrev} and numerical simulations \cite{CSsimuls}
tell us that the network of cosmic strings will take on
a ``scaling solution" in which the average quantities describing the
network are invariant in time if measured in Hubble length
$H^{-1}(t)$, where $H$ is the expansion rate of space.

In this paper we will compute the angular power spectrum at
fixed redshift produced by the distribution of cosmic strings emitting 21~cm radiation. 
Specifically, we are interested in the shape and overall amplitude of the signal.
The paper is organized as follows: We begin in Section \ref{sec:wake-21cm} with a brief 
review of cosmic string cosmology and observables of cosmic string wakes in 
21~cm experiments. 
In Section \ref{sec:cs-wakes}, 
we review the cosmic string toy model of \cite{Periv} which we use, 
and construct the cosmic string wake profile and number statistics. In
Section \ref{sec:power-spectrum}, we calculate the 21~cm power spectrum
for these cosmic string wakes.  Section \ref{sec:power-spectrum}
contains the main result of the paper. Readers who are familiar with
cosmic string wakes or 21~cm observations can skip Section
\ref{sec:wake-21cm} or Section \ref{sec:cs-wakes} and jump directly
to Section \ref{sec:power-spectrum}. In Section~\ref{sec:conclusion} we 
present our conclusions and put our work in context. 

\begin{figure}
\includegraphics[height=4.7cm]{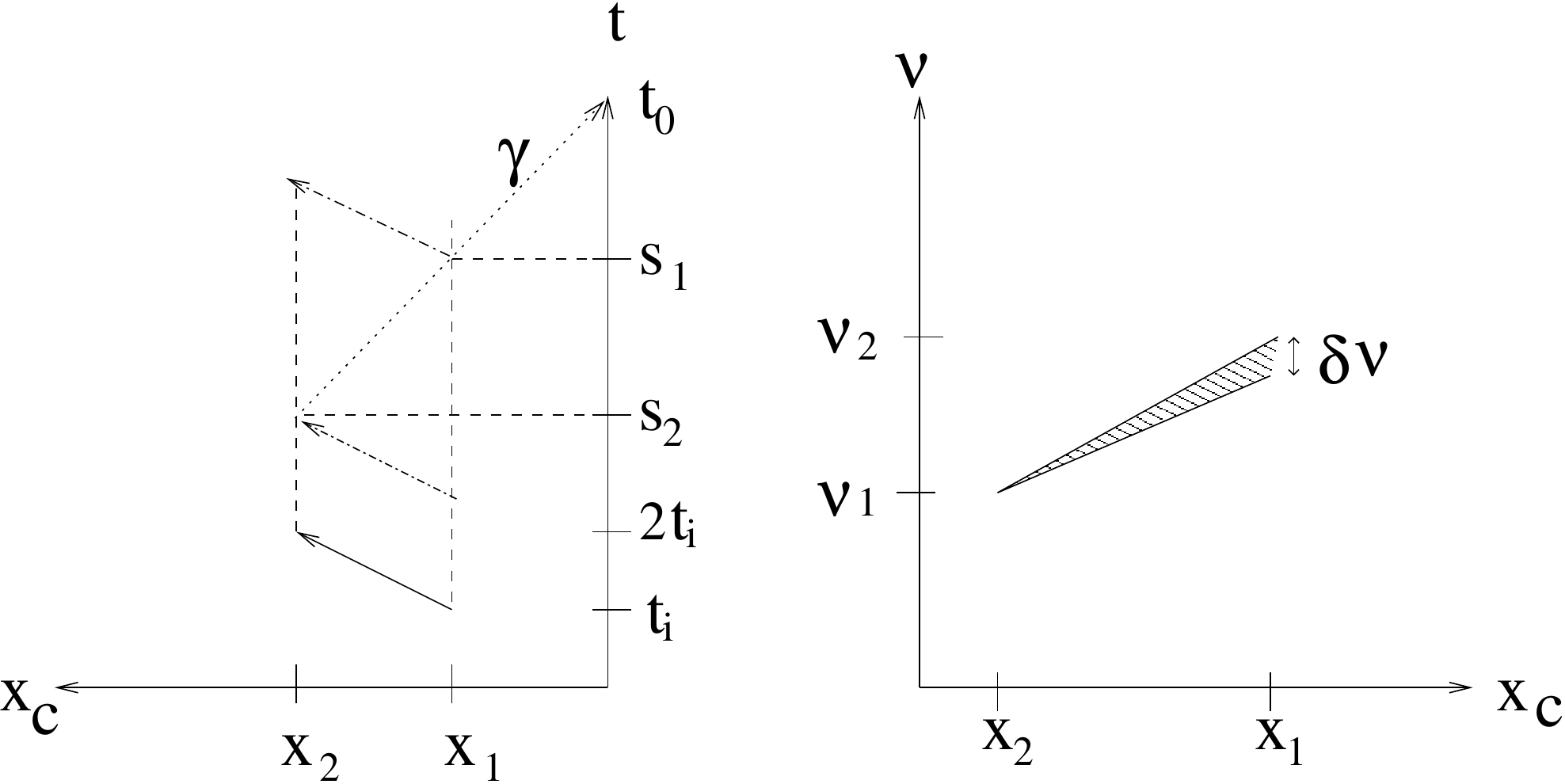}
\caption{Geometry of the 21~cm signal of a cosmic string wake. The
left panel is a sketch of the geometry of the wake in space-time -
vertical axis denoting time, the horizontal axis denoting one direction of
space. The string segment producing the wake is born at time $t_i$ 
and travels in the direction of the arrow, ending at the position 
$x_2$ at the time $2 t_i$. The past light cone of the observer
at the time $t_0$ intersects the tip of the wake at the time $s_2$,
the back of the wake at time $s_1$. These times are in general
different. Hence, the 21~cm radiation from different parts of the
wake is observed at different red-shifts. The resulting angle-redshift
signal of the string wake shown in the left panel is illustrated
in the right panel, where the horizontal axis is the same spatial
coordinate as in the left panel, but the vertical axis is the
redshift of the 21~cm radiation signal. The wedge in 21~cm has vanishing
thickness at the tip of the wedge, and thickness given by $\delta \nu$
at the back side.} \label{fig:4}
\end{figure}

\section{Cosmology of Cosmic Strings and Their Wake's Brightness Temperature}
\label{sec:wake-21cm}

Cosmic strings are lines \footnote{These lines are in fact thin tubes with 
thickness given by a microphysical scale.} in space with trapped energy. 
It is the gravitational effects of this energy which leads to the
role which the strings play in the formation of cosmological structure, as
first pointed out in \cite{ZelVil}.
As relativistic objects they have tension equal to the mass per unit length and velocity 
$v_s$ of the order $c$. Their relativistic nature is also responsible for the
characteristic string signatures in cosmological observations - the fact that space 
perpendicular to a long straight string is
not flat, but conical with deficit angle $\alpha \, = \, 8 \pi G \mu$ \cite{deficit}.
This deficit angle leads to gravitational lensing of microwave radiation
about the string which in turn leads to a line
discontinuity in the temperature on the CMB \cite{KS}. These line
discontinuities can be searched for \cite{ABB} using edge detection algorithms
such as the Canny algorithm \cite{Canny}. Maps with good angular resolution
such as those that are being produced by the South Pole Telescope (SPT) \cite{SPT}
and Altacama Cosmology Telescope (ACT) \cite{ACT} are ideally suited to search
for cosmic strings.

A long straight cosmic string segment moving through space will lead to
a wedge-shaped overdense region in its wake, a so-called ``cosmic string
wake" \cite{wake}. To understand what a wake is let us look at the physics 
from the point of view of an observer on the string.
This observer will see matter streaming by on both sides. But since space
is conical, the matter acquires a velocity $\delta v$ given by
\be
\delta v \, = \, 4 \pi G \mu v_s \gamma(v_s)
\ee
towards the plane behind the string. This leads to the formation of a
wedge of twice the background matter density behind the string.
The wake, in turn, will grow in thickness by gravitational
accretion, as studied e.g. in \cite{wakegrowth} using the Zel'dovich
approximation. 
If a cosmic string
is laid down at a redshift $z_i$, then the mean width of the wake
(defined as the overdense region which has collapsed and undergone
shock heating) at redshift $z_e$ will be 
\be \label{width}
w(t_e) \, = \, \frac{6 \pi}{5} G \mu v_s \gamma(v_s) 
t_i  \bigl[ \frac{\bigl( z_i + 1 \bigr)}{\bigl( z_e + 1 \bigr)} \bigr]^{2} \, .
\ee

The extra baryon density in the wake will lead to extra 21~cm absorption or
emission.  
If the 21~cm radiation is emitted from a wake at redshift $z_e$ its brightness temperature 
is given by
\be \label{eq:deltaTb}
\delta T_b(z_e) \ = \, [70~{\rm mK}]\frac{x_c}{1+x_c}\left(1-\frac{T_\gamma}{T_K}\right){(1+z_e)^{1/2}\over2\sin^2\theta}~.
\ee
Here $T_\gamma$ is the CMB temperature, $T_K$ is the wake gas kinetic temperature, 
$\theta$ is the angle of the 21~cm ray with respect to the vertical to the wake, and 
$x_c$ are the collision coefficients whose values can be obtained from the tables 
listed in \cite{xc}. Here and throughout we take the cosmological parameters to be  
$H_0=73~{\rm km~s}^{-1}~{\rm Mpc}^{-1}$, $\Omega_b=0.0425$, $\Omega_m=0.26$. We work with a matter dominated universe for $z\le3000$ with the age of the universe $t_0=4.3\times10^{17}$~s.

Note the $T_K$ and the $\theta$ dependence  in Eq.~\ref{eq:deltaTb}. The string tension 
and the string's speed affect the temperature $T_K$ of the neutral hydrogen in the wake, 
while the string's direction of motion determines $\theta$. We simplify our calculations 
below by working with the brightness temperature of a string oriented so that 
$2\sin^2\theta=1$ and $(v_s \gamma(v_s))^2=1/3$. Thus the only string parameter 
that will determine the brightness temperature will be the string tension $G \mu$.

Whether the string signal is an emission or absorption signal (positive or negative 
brightness temperature, respectively)
depends on the ratio between the wake temperature of neutral hydrogen and the 
temperature of the microwave photons. The former is larger at lower redshifts,
the latter is larger at higher redshifts. At redshifts below $z=20$, reionization effects 
become important so we consider 21~cm radiation emitted at $z_e=20$ from a string 
formed at the redshift $z_i + 1 = 3000$ 
corresponding to the time $t_{eq}$
of equal matter and radiation. 
We found \cite{us}
that the critical value of $G \mu$ at which the transition from absorption
to emission occurs is $(G \mu)_6 \, \simeq \, 0.23 ,$
where $(G\mu)_6$ is the string tension in units of $10^{-6}$.

\section{Cosmic string wake networks}
\label{sec:cs-wakes}

\subsection{Scaling solution}

The scaling solution is characterized by a random walk-like network
of ``infinite" strings (length larger than the Hubble radius) with step length
comparable to the Hubble radius, plus a distribution of string loops with
radii $R$ smaller than the Hubble radius. According to numerical 
simulations \cite{CSsimuls}, the long strings dominate. For our study,
we will make use of an analytical toy model for the distribution of strings
first introduced in \cite{Periv}: we divide the relevant time interval (in
particular the time interval between $t_{eq}$ and the present time $t_0$)
into Hubble time steps. At each step, space is divided up into
boxes of Hubble size at the corresponding local times. 
In each time interval we lay down $N_H$ straight
cosmic string segments per Hubble volume. According to numerical simulations the value of $N_H$ ranges from
$1$ to $10$. Each string segment laid down at time $t$ 
has a physical length at that time which is
given by $c_1 t$ (where $c_1$ is a constant which is expected
to be slightly smaller than $1$, and whose value is expected to decrease as
$N_H$ increases) and depth given by $v_s \gamma(v_s) t$,
where $v_s$ is the velocity of the string in the plane perpendicular to
its tangent vector, and $\gamma(v_s)$ is the associated relativistic gamma
factor. The value of $v_s$ is expected to be of order one in units where the speed
of light $c$ is set to one. 

The earliest Hubble 
time interval which we consider is at the time of equal matter and radiation,
and the corresponding size of the spatial boxes at that time is
$t_1 = t(z_{\rm eq})$. The wakes produced by these strings are in fact both
the most numerous (since a fixed comoving volume contains the
largest number of wakes of that size) and the thickest (since the dark matter
fluctuations have had the longest time to grow. 
The baryons are tightly 
coupled to radiation until the time $t_{\rm rec}$ of recombination but then
fall into the potential wells created by the dark matter.
String wakes laid down 
before $t_{\rm eq}$ will have a thickness which scales as $(z(t_i) + 1)^{-1/2}$
since the dark matter perturbations cannot grow while the dark matter
forms the subdominant component of matter.  

Later time steps are centered at the values $t_2 = 2t(z_{\rm eq})$, 
$\ldots$, $t_m = 2^{m-1} t(z_{\rm eq})$ respectively. Thus the total
number of these Hubble time steps is
\be
 M_{\rm tot} \, = \, 
{\rm Int}\left[\log_2\left(\frac{t_0+t(z_{\rm eq})}{t(z_{\rm eq})}\right)\right]
 \, = \, {\rm Int}\left\{\log_2\left[1+\left({z_{\rm eq}+1}\right)^{3/2}\right]\right\}
  ~,
\ee
where Int$(x)$ is the integer part of $x$. Taking $z_{\rm eq}\simeq 3000$, we
have $M_{\rm tot}=17$. However, as we discuss later in Section
\ref{sec:bound-hubble-steps}, not all of these Hubble steps (at which the wakes are
laid down) contribute to the 21~cm emission power spectrum. 

At each Hubble step, the cosmic strings are taken to be ``created" at the
beginning of the Hubble step and ``annihilated" at the end of the step. In
reality, the string network is a complicated dynamical distribution. Since the
strings are moving relativistically, intercommutation events of two strings
are frequent. For a fixed string, it lasts typically about one Hubble time step
until this string intersects another one. Such an intersection will lead to
a coarsening of the network of long strings via the  production of cosmic
string loops and the consequent straightening of the remaining long string
network. Hence, we are modelling the distribution of the long string network
in any Hubble time step as a collection of straight string segments which
result from an initial intercommutation event at the time of
``creation'', and undergo a next intercommutation at the time of 
``annihilation''. Since the intercommutation events are uncorrelated
in space, it is a good approximation to consider the centers and directions
of motion of the string segments to be randomly distributed, and
uncorrelated from one time step to the next. Similarly,
we expect a spread in the distribution of the velocities $v_s$. 

The cosmic string wakes are produced by the motion of the strings through 
space. Once produced, the wakes will be approximately static in comoving
coordinates, except for the growth in thickness due to gravitational
accretion of the surrounding gas. We will
consider the shape profile and number statistics of these wakes in the
following subsections. 

\subsection{Wake profile}
\label{sec:wake-profile}

A cosmic string segment laid down at time $t_i$ will generate a wake whose
physical dimensions at that time are
\be
  c_1 t_i ~\times~ t_i v_s\gamma_s ~\times~ 4\pi G\mu t_i v_s\gamma_s \, .
\ee
The dimensions $c_1 t_i$ and $t_i v_s\gamma_s$ span the two length 
dimensions of the wake and are independent of the string tension. They are 
both of the order of the instantaneous Hubble radius. The third dimension 
is the width of the wake, which is much smaller than the length because it
is suppressed by the small parameter $G \mu$.

These three dimensions evolve with time.  Firstly, the two length dimensions
are comoving in space. Thus at a later time $t_e$, their physical size is
\be
  c_1 t_i \left(\frac{z_i + 1}{z_e + 1}\right) ~\times~ v_s\gamma_s
  t_i \left(\frac{z_i + 1}{z_e + 1}\right)  \, ,
\ee
where $z_i = z(t_i)$ and $z_e = z(t_e)$.
For simplicity, in some of the later calculations, we shall use $l(z_i, z_e)$ to denote
the length of the wake at time $t_e$ if the time of formation is $t_i$, 
ignoring the order one difference between these two dimensions.

The width of the wake will grow in time. According to  linear perturbation
theory if follows that the comoving width dimension grows linearly with the scale 
factor. We are interested in the width of the region which has collapsed and
shock heated. The determination of this region can be done using the
Zel'dovich approximation \cite{Zel} and gives a result which is smaller than
the naive linear perturbation theory result by approximately a factor of $4$ \cite{us}.
The final result is given in (\ref{width}).

\subsection{Wake Number Statistics: Method 1}
\label{sec:wake-numb-stat1}

For later use, we will calculate the number density of cosmic string wakes on a
fixed redshift hypersurface in two ways, making use of two different approximations. 
Both calculations agree to within a factor of order one. 

We shall use $S^2$ to denote the hypersurface with
fixed redshift $z_e$ from which the 21~cm signals are emitted.
The comoving radius $r$ of this sphere can be calculated as
\begin{equation}
 r(t_e) \, = \, \int_{t_e}^{t_0}\frac{dt}{a(t)}=\frac{3}{a_0}t_0^{2/3}(t_0^{1/3}-t_e^{1/3})~,
\end{equation}
where $a_0$ is the value of the scale factor at the present time $t_0$ and can
be set to $1$ without loss of generality.

At the time $t_m$ when wakes in the m'th Hubble time interval are laid down,
the physical radius of the above $S^2$ is
\begin{equation}
 R_m \, = \, a(t_m) r(t_e) \, = \, 3 t_m^{2/3}(t_0^{1/3}-t_e^{1/3})~.
\end{equation}
Except at the final Hubble time steps close to the present time, this value
of $R_m$ is much greater than the Hubble scale $1/H(t_m)=3t_m/2$, which
is also the typical length of straight string segments. Thus, for our
counting procedure, the string segments
can be treated as point particles.
Note that the length of the string wake is of the same order as the Hubble
  scale. This breaks the particle approximation which is made here.
  However, as we show in the next section, using a full string word sheet
  calculation, the simple and intuitive approximation made here leads
  to the correct result, up to a constant of order one.

The number of strings that pass through the $S^2$ per unit area per unit time
is
\begin{equation}
  n_s \times v_s \times \cos\theta_s~,
\end{equation}
where $n_s,~v_s,~\theta_s$ denote the number
density, averaged velocity and direction of motion of the cosmic strings,
respectively. Thus, the number of strings that crosses the the $S^2$ is
\begin{equation}\label{eq:ns21}
  N_{S^2} \, = \, n_s\times v_s\times \cos\theta_s\times 4\pi R_m^2\times 1/H(t_m)~.
\end{equation}
Since the cosmic string network achieves a scaling solution, 
it is more convenient to use number $N_H$ of strings
per Hubble volume instead of the number density of strings. Thus, we have
\begin{align}
 N_{S^2} \, =  \, N_H\times v_s\times \cos\theta_s\times 4\pi R_m^2\times H(t_m)^2~.
\end{align}
This $N_{S^2}$ is also the number of wakes on the sphere where the 21~cm signal
with redshift $z_e$ were emitted. 

The two dimensional number density of cosmic string wakes at the emission time
$t_e$ can be calculated as
\be \label{n2D}
  n_{2D} \, = \, \frac{N_{S^2}}{4\pi a(t_e)^2 r^2} \,
= \, \frac{4N_Hv_s\cos\theta_s}{9t_m^2}
  \left(\frac{z_e+1}{z_m+1}\right)^2~.
\ee
For our examples below we have averaged the direction speed of motion
such that $\cos^2\theta_s=1/2$ and $v_s=1/2$, and take $N_H=10$,
$t_m=2.6 \times 10^{12}$ s for the first Hubble step ($m=1$) corresponding to $z=3000$. 

\subsection{Wake Number Statistics: Method 2, a Worldsheet Calculation}
\label{sec:wake-numb-stat2}

We calculate the probability that a string world sheet intersects the $S^2$. The
string world sheet is created at $t_m$, and destroyed one Hubble time after
that, with string length $l_s$. Note that the length scale
$l_s$ is of order $t_m$. Thus the current method should be more
precise than the previous one.

For simplicity, we approximate the string world sheet as circular. The
rectangular world sheet could also be calculable, but with the need of 
taking into account considerably more effects. We use $r_{\rm ws}$ to 
denote the radius of the circular world
sheet. To best mimic a rectangular world sheet, we take the area of the world
sheet to be unchanged while deforming its shape. Thus we have
\begin{equation} \label{size}
  {\pi}r_{\rm ws}^2 \, = \, c_1 t_m \times t_m v_s\gamma_s ~,
  \qquad
  r_{\rm ws} \, = \, t_m \sqrt{c_1 v_s\gamma_s/\pi} ~.
\end{equation}

\begin{figure}
\centering
\includegraphics[width=0.5\textwidth]{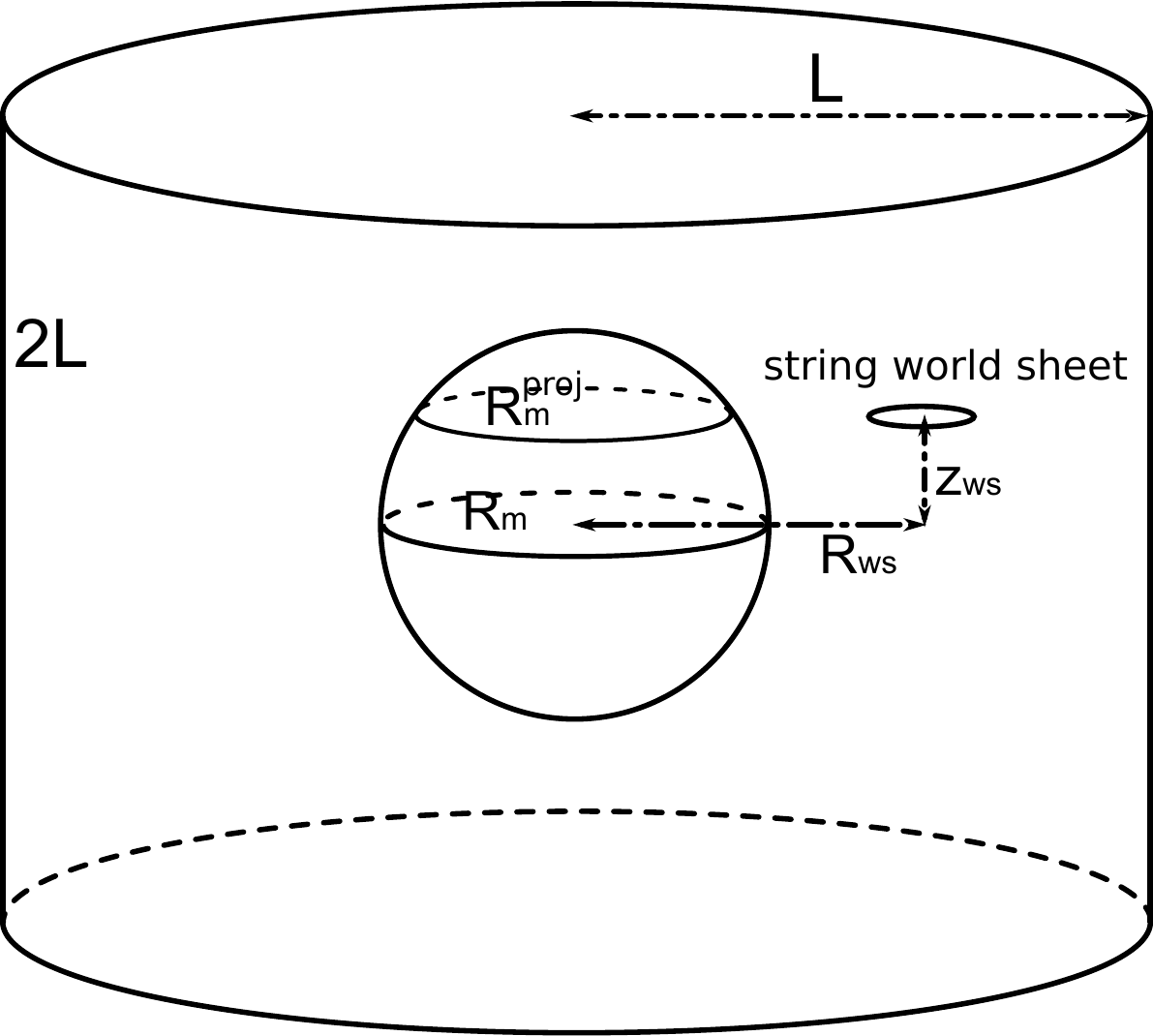}
\caption{\label{fig:ns22} Illustration of how the number of relevant
wakes is computed in Section~\ref{sec:wake-numb-stat2}. 
Here the $z_{\rm sw}$ direction is chosen to be
  perpendicular to the string world sheet. And in the figure we have rotated the
  $z$ direction to be the vertical direction only for the purpose of convenience.}
\end{figure}

Consider one single circular cosmic string world sheet, as shown in Figure
\ref{fig:ns22}.  Without loss of generality, we consider the world sheet inside a
cylindrical volume. The cylinder has volume $2\pi L^3$. Here we take $L > R$, 
where $R$ is the largest of the scales appearing in (\ref{size}). The
probability for the string world sheet (which is inside this clinder) to
intersect the $S^2$ can be calculated as
\begin{equation}
  \frac{1}{2\pi L^3}\int_{-R_m}^{R_m} dz_{\rm ws} \int_0^{L}dR_{\rm ws}\int_0^{2\pi}
  R_{\rm ws}d\theta_{\rm ws}\times \Theta(R_{\rm ws}-R_m^{\rm proj}+r_{\rm
    ws})\Theta(R_m^{\rm proj}+r_{\rm ws}-R_{\rm ws})~.
\end{equation}
where $\Theta(x)$ is the Heaviside step function, and 
$R_m^{\rm proj}\equiv \sqrt{R_m^2-z_{\rm ws}^2}$. Keeping in mind that
$R_m\gg r_{ws}$, the above integration yields
\begin{equation}
  ({\rm Probability}) \, \simeq \, \frac{\pi R_m^2 r_{\rm ws}}{L^3}~.
\end{equation}
Thus, for $N_{\rm ws}$ world sheets, the number of intersections is 
$N_{\rm ws} \pi R_m^2 r_{\rm ws} /L^3$. The number of world sheets 
$N_{\rm ws}$ is related to the
number density of world sheets (which equals to the number density of strings)
by $N_{\rm ws} = 2\pi L^3 n_s$. Thus, finally, we have
\begin{equation}
  N_{S^2} \, = \, 2\pi^2 n_sR_m^2 r_{\rm ws} =(4\pi n_s v_s R_m^2 H_m^{-1}) 
   \left({\sqrt{\pi c_1 v_s\gamma_s}\over3v_s\cos\theta_s}\right)
  \, .
\end{equation}
When the factor $\sqrt{\pi c_1 v_s\gamma_s}/(3v_s\cos\theta_s)$ is of order unity, this result is in agreement with the result
\eqref{eq:ns21}. 

\subsection{Cutting Off the Hubble Steps}
\label{sec:bound-hubble-steps}

For the calculations in~\cite{us} to hold the wake must undergo shock heating. 
This occurs when the kinetic temperature inside the
wake is at least 2.5 times greater than the background gas
temperature (the temperature of the matter particles outside of the wake,
not the photon temperature $T_{\gamma}$). 
Shock heating is more likely
to have occurred for strings laid down earlier. 
Below we determine the upper cutoff time for shock heating to have occurred.

Note that the later the wake is laid down, the lower the kinetic
temperature. This leads to a lower bound on the redshift $z_m$, after which
the cosmic string wakes will not have undergone shock heating. The kinetic temperature 
for the gas inside the wake is \cite{us}
\be
  T_K \, = \, [20{\rm K}](G\mu)_6^2 (v_s\gamma_s)^2  \frac{z_m+1}{z_e+1} \, .
\ee
Until reionization, the background gas is cooling adiabatically below redshifts of 
$z \approx 150$ as:
\be
  T_g \, = \, 0.02{\rm K}(1+z_e)^2 \, .
\ee
The requirement $T_K > 2.5 T_g$ leads to 
\be \label{eq:cut-off-z}
  z_m+1 \, > \, \frac{(1+z_e)^3}{400(G\mu)_6^2(v_s\gamma_s)^2} \, .
\ee
Only the Hubble steps with $z_m$ satisfying Eq. (\ref{eq:cut-off-z}) lead to wakes where shock heating occurs.

This requirement that only shock heated wakes contribute significantly leads to 
a maximal number of Hubble steps
in the toy model. The Hubble steps that should be taken into account are
$m=1,\ldots, M$, where
\be
  M \, = \, {\rm Int}\left[\log_2\left(\frac{t(z_m^{\rm min}) + t(z_m^{\rm
          max})}{t(z_m^{\rm max})}\right)\right]
  \ = \, {\rm Int}\left\{\log_2\left[1+\left(\frac{z_m^{\rm max}+1}{z_m^{\rm min}+1}\right)^{3/2}\right]\right\}
  ~,
\ee
where Int$(x)$ denotes the integer part of $x$.
Later wakes which have not 
undergone shock heating are thinner (see Eq.~\ref{width}) and less numerous (see Eq.~\ref{n2D}) than the earlier ones. In 
addition, they are not gravitationally bound due to free streaming. For both of these 
reasons, their contribution to our power spectrum is negligible.

\begin{figure}[htpb]
\centering
\includegraphics[width=0.8\textwidth]{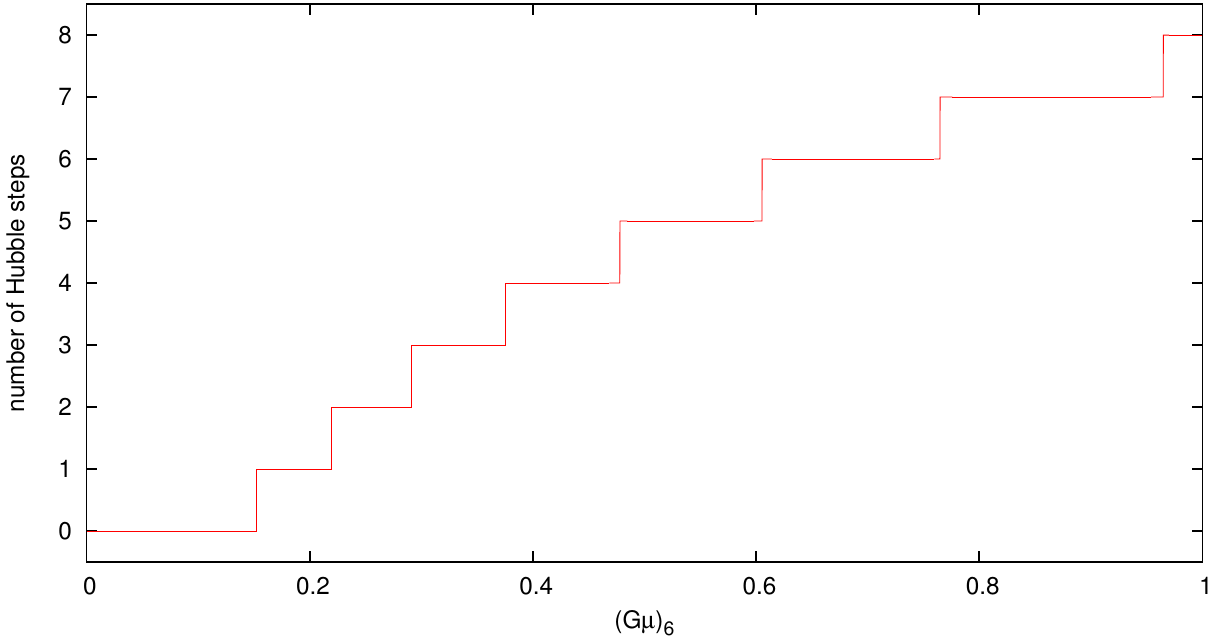}
\caption{\label{fig:steps} The number of Hubble steps, for a given string tension, during which a wake emits 21~cm radiation. The parameters used are $z_e=20$, $z_m^{\rm max}=3000$
  and $v_s^2\gamma_s^2=1/3$. }
\end{figure}

For example, when we take the 21~cm emission redshift to be $z_e = 20$, and set the
maximal redshift of cosmic string wake laid down time to be the matter radiation
equality time $z_m^{\rm max}\simeq 3000$, then the number of Hubble steps which lead to important cosmic 
string wake 21~cm emission is as plotted in Fig. \ref{fig:steps}.

\section{Power spectrum}
\label{sec:power-spectrum}

We will compute
the angular power spectrum of the 21~cm brightness temperature at a
fixed redshift with a string tension larger than the critical value for emission. 
A few words of warning are in order. Firstly,
for a Gaussian distribution the two point function contains all the information
about the distribution. However, in the case of cosmic strings, the signal is
very non-Gaussian. Hence, the two point correlation function will miss
important information. Secondly, one of the advantages of 21~cm surveys
compared to other measures (e.g. CMB temperature maps) is that the
maps are three dimensional and thus can contain
much more information. In restricting attention to a fixed redshift, we
are not exploiting this advantage. Nevertheless, in order
to compare the 21~cm signal of cosmic strings with that of other sources,
it is useful to compute the angular correlation function. We plan to
extend our analysis and compute the three dimensional correlation
function in upcoming work.

We are performing our calculation in the framework of the toy model
distribution of long strings of \cite{Periv} reviewed earlier. In this
analysis, the strings at different Hubble time intervals are statistically
independent. Hence, we can first consider a fixed string formation time
Hubble interval and sum the results over all redshift intervals which
make a contribution. In the following subsection, we therefore consider
the contribution from strings in a fixed single Hubble string formation
time step.

\subsection{A single Hubble step}
\label{sec:single-hubble-step}

First we calculate the two point correlation function 
$\langle\delta T_b({\bf 0})\delta T_b({\bf x})\rangle$
The brackets indicate an ensemble
average. We assume the validity of the ergodic hypothesis so that ensemble 
averages can be computed by volume averages. 
We shall use the flat
sky approximation because we are interested in scales of order ten degrees 
on the sky or smaller. 
The position $\bf x$ in the two dimensional sky is the 2-d angular coordinate times the 
radial distance of the fixed redshift surface $z_e$ under consideration.

\begin{figure}[htpb]
\centering
\includegraphics[width=0.8\textwidth]{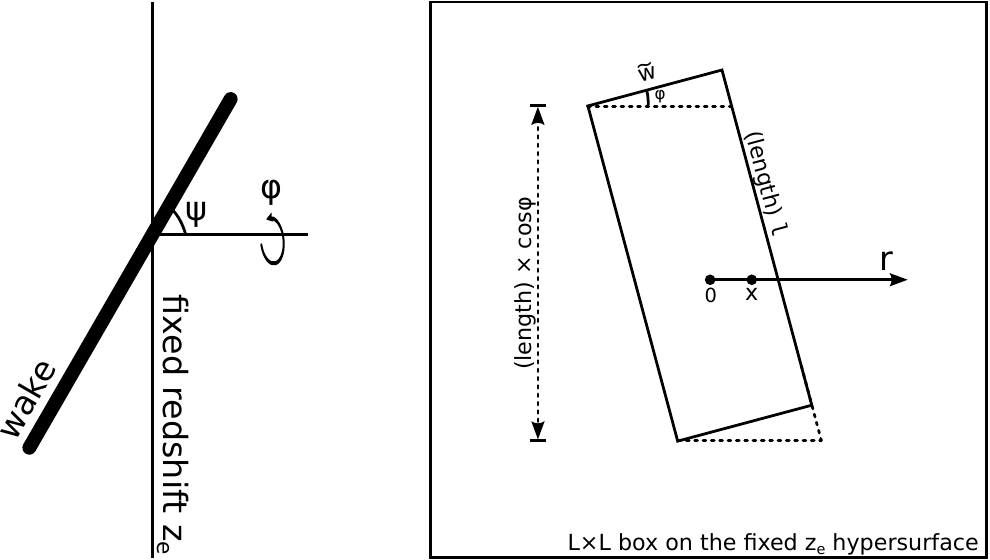}
\caption{\label{fig:powerCal} This figure illustrates the cosmic string wake
  profile. Both panels represent cross sections through space at the time
$t_e$. The ``fixed redshift hypersurface" is the intersection of our past
light cone with this space. The left panel introduces the angles which describe
the relative geometry between the wake and the
fixed redshift hypersurface, on which the 21~cm signal is emitted. The right
panel  shows the projection of the wake onto the fixed redshift surface. Here the
 ``width'' in the figure  is $\tilde w$, the projected width of the wake.}
\end{figure}

Whereas in the three dimensional redshift-angle space a wake is long
in two dimensions and narrow in one, the intersection of the wake with
the fixed redshift surface produces an object which is long in
only one dimension and typically narrow in the second. We will call
the size of this second dimension the ``projected width". As
illustrated in Fig. \ref{fig:powerCal}, the projected width $\tilde w$
of the wake can be written as
\be
  \tilde w \, = \,  
  \left\{
    \begin{array}{rcl}
      w / \cos\psi &\mbox{for} &
      0\leq \psi<\pi/2 - \arcsin (w/l) \\
      l &\mbox{for} & \pi/2 - \arcsin (w/l) \leq \psi\leq \pi/2
    \end{array}
  \right. \, ,
\ee
where $\psi$ is the angle of the wake relative to the fixed redshift surface.

Since the positions and orientations of different string wakes at a fixed Hubble
time step are taken to be random, the computation reduces to the computation
of the ensemble average of the contribution of a single string wake.
We will consider the ensemble average of the contribution of a single string in a
box of dimensions $L\times L$ , where $L$ is lager than the length of the wake we
are interested in. Since the dominant contribution to the 21~cm signal comes from
strings created at about $t_{\rm eq}$, we do not need $L$ to correspond to a large
angular scale. As in the studies of position space signals of cosmic strings
to CMB temperature and polarization maps, an angular extent of $10^{\circ}$
is more than enough (larger areas will obviously lead to reduced statistical
error bars). Returning to the study of the effect of a single string wake,
the geometry is illustrated in Fig. \ref{fig:powerCal}. Here we 
assume that the cosmic string wake intersects the surface
of fixed redshift $z_e$. As discussed in Section \ref{sec:wake-profile},
the number density of such string wakes is $n_{2D}$. We now study
the contribution of this string to the correlation function.

When both points $0$ and ${\bf x}$ are on the wake, the product 
$\delta T_b({\bf 0})\delta T_b({\bf x})$
contributes $\overline{\delta T_b}^2$ to the ensemble average. Otherwise, when either
points $0$ or $\bf x$ are not on the wake, the product is zero. In this way, the
calculation of the ensemble average is reduced to the calculation of probability
that both points $0$ and $\bf x$ are on the wake. We have
\be \label{cor1}
  \langle \delta T_b({\bf 0})\delta T_b({\bf x}) \rangle \, = \, \overline{\delta T_b}^2 \times \mbox{(Probability)}~.
\ee
There are probabilities for both the horizontal and vertical coordinates of the 
points $0$ and $\bf x$ to be on the wake. When $0<\phi<\arctan (l/\tilde w)$
($\phi$ is labeled in Fig. \ref{fig:powerCal}), the probability we
calculate is a product of these two probabilities:
\be \label{prob1}
  \mbox{(Probability)} \, = \,  
  \left\{
    \begin{array}{rcl}
      \frac{\tilde w/\cos\phi - r}{L} \times \frac{l\cos\phi}{L} &\mbox{for} &
      r<\tilde w/\cos\phi \\
      0 &\mbox{for} & r\geq \tilde w/\cos\phi
    \end{array}
  \right. ~,
\ee
where $r$ is the magnitude of ${\bf x}$ and 
$\tilde w \equiv w/\cos\psi$ is the projected width of the wake onto the
fixed redshift hypersurface. The angles $\psi$ and $\phi$ are shown in
Fig. \ref{fig:powerCal}. They should be integrated over when performing
the ensemble average.

On the other hand, there is the other region 
$\arctan(l/\tilde w)<\phi<\pi/2$, which is small on 
small scales but will give a comparable
contribution on large scales (scales of order the length of the wake). In
this region, the probability that both points 0 and $\bf x$ are on the
wake can be written as
\be \label{prob2}
  \mbox{(Probability)} \, = \,  
  \left\{
    \begin{array}{rcl}
      \frac{l\sin\phi - r}{L} \times \frac{\tilde w/\sin\phi}{L} &\mbox{for} &
      r<l\sin\phi \\
      0 &\mbox{for} & r\geq l\sin\phi
    \end{array}
  \right. ~.
\ee

Based on rotational symmetry of the ensemble, the correlation function
will only depend on $r$, the magnitude of ${\bf x}$. 
For a single wake, the considerations of (\ref{cor1}), (\ref{prob1}) and (\ref{prob2}) yield
\begin{align}\label{eq:position-space-corr}
  \langle \delta T_b(0)\delta T_b(r) \rangle &
  = \frac{2\overline{\delta T_b}^2}{\pi L^2}
  \int_0^{\pi/2}d\psi \int_0^{\arctan
    (l/\tilde w)}d\phi ~ l\sin\psi \left(
    \tilde w-r\cos\phi \right)\Theta(\tilde w/\cos\phi -r)
  \nonumber \\ &
  + \frac{2\overline{\delta T_b}^2}{\pi L^2}\int_{0}^{\pi/2}d\psi\int_{\arctan
    (l/\tilde w)}^{\pi/2}d\phi ~ \tilde w\sin\psi  (l-r/\sin\phi)
  \Theta(l\sin\phi-r)~.
\end{align}
where $\Theta(x)$ is the Heaviside step function. Note that
we have put a normalization factor $2/\pi$ in front of the integration, because
we are averaging over 1/8 of the solid angle by restricting $\psi$ and $\phi$
to both run from 0 to $\pi/2$. 

The angular power spectrum of zero mean fluctuations 
is obtained by taking the Fourier transform of
the above correlation function and subtracting  the square of the brightness 
temperature times $(2\pi)^2 \delta^2_D(\vec k)$ for the zero mode.
Hence for $k\ne0$ the 
contribution of a single string wake can be written as
\be \label{eq:power-def}
  P_1(k) \, =\, \int_0^\infty dr \int_0^{2\pi} d\theta ~ re^{-ikr\cos\theta} \langle \delta T_b(0)\delta T_b(r) \rangle ~,
\ee
where $r,\theta$ are the polar coordinates of {\bf{x}}.
One can insert Eq. (\ref{eq:position-space-corr}) into Eq. (\ref{eq:power-def}),
integrate out $\theta$, $r$, $\phi$ and $\psi$ to obtain the power
spectrum on length scales much smaller than the wake length $l$. In this regime, the power
spectrum can be well approximated by
\be
  P_1(k) \, = \, \frac{4l\overline{\delta T_b}^2}{L^2k^3}
  \left[
    1-\cos(wk)-wk{\rm Si}(wk)+wk{\rm Si}(lk)
  \right]~,
\ee
where ${\rm Si}(z)$ is the sin integral defined as ${\rm Si}(z)\equiv \int_0^z dt~
(\sin t)/t$. On the other hand, 
on scales larger than the wake length $l$, the power spectrum $P_1(k)$ is independent 
on $k$. This can be seen by taking the limit $k\rightarrow 0$ in 
Equation (\ref{eq:power-def}). The full power spectrum can be calculated numerically.

\begin{figure}[htpb]
\centering
\includegraphics[width=0.8\textwidth]{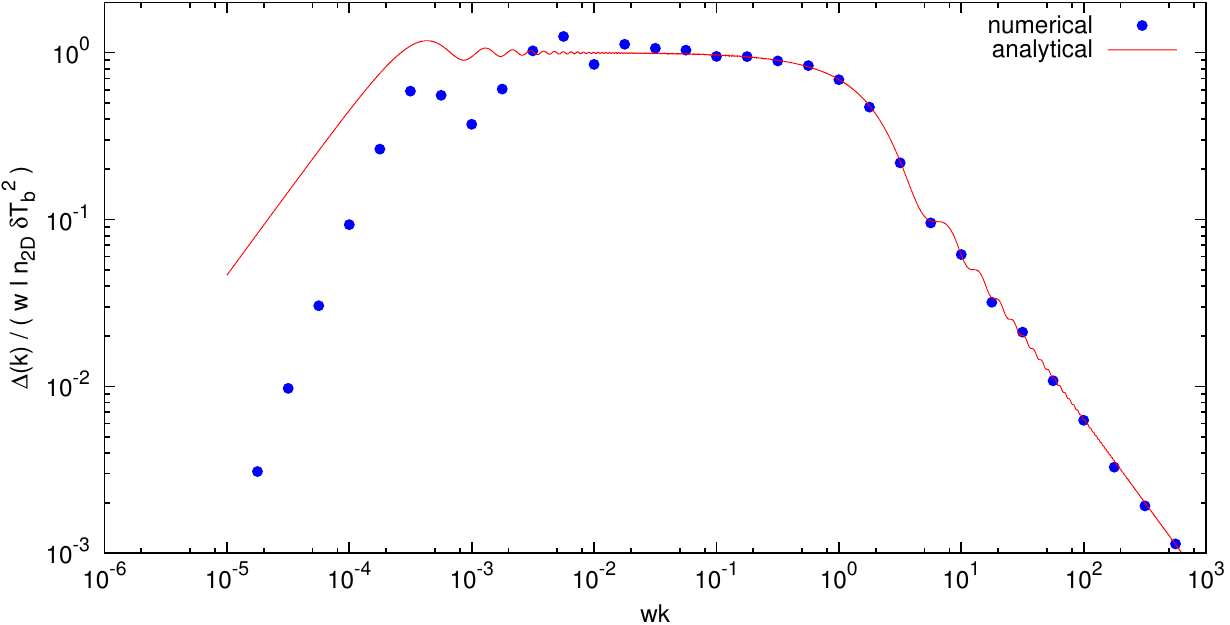}
\caption{\label{fig:powerFig} The dimensionless power spectrum $\Delta(k)$. We plot
  ${\Delta(k)}/{(wl n_{2D} \overline{\delta T_b}^2)}$ as a function of $wk$. In the plot we
take $l/w$ to be $9000$. This value is calculated using $z_e=20$, $z_m=3000$,
$(G\mu)_6=0.2$, $v_s=1/2$, and $\cos^2\theta_s=1/2$.
}
\end{figure}

Now we will consider multiple cosmic string wakes which are laid down within
the same Hubble time step. Up to a factor of order one, these cosmic string wakes 
have the same dimensions as calculated in Section \ref{sec:wake-profile}. If we assume
these wakes do not have any cross correlations between each other, then
we simply have to multiply the result for a single string wake by the number of
string wakes in the observation area, which is $L^2 n_{2D}$. Thus, the
result for multiple string wakes (single Hubble time step) in the region $kl>>1$ becomes
\be
  P(k) \, = \, \frac{4l n_{2D} \overline{\delta T_b}^2}{k^3}
  \left[
    1-\cos(wk)-wk{\rm Si}(wk)+wk{\rm Si}(lk)
  \right] \, .
\ee
On scales much smaller than the wake length $l$, the dimensionless power spectrum 
\footnote{Here dimensionless means that there is no $k$ dimension. The power 
spectrum still has dimension of temperature squared.}
can be written as
\be \label{eq:Delta}
  \Delta(k) \, \equiv \, \frac{k^2}{2\pi}P(k) \, = \, 
  \frac{2l n_{2D} \overline{\delta T_b}^2}{\pi k}
  \left[
    1-\cos(wk)-wk{\rm Si}(wk)+wk{\rm Si}(lk)
  \right] ~.
\ee
Eq. (\ref{eq:Delta}) and the corresponding numerical result 
(shown in Fig. \ref{fig:powerFig}) are the main result of our paper.

Since, as mentioned earlier, $P(k)$ is constant in the limit of small values
of $k$, $\Delta(k)$ scales $k^{2}$ for small $k$. 
The scaling for large values of $k$ is $k^{-1}$, which can be obtained by expanding  
Eq. (\ref{eq:Delta}) in a power series. 

It is also convenient to use $wk$ as the argument and rewrite the dimensionless
power spectrum (for $kl>1$) as
\be \label{eq:Delta1}
  \frac{\Delta(k)}{wl n_{2D} \overline{\delta T_b}^2} \, = \, \frac{2}{\pi wk}
  \left[
    1-\cos(wk)-wk{\rm Si}(wk)+wk{\rm Si}\left(\frac{l}{w} \times wk \right)
  \right]~.
\ee
Eq. (\ref{eq:Delta1}) is plotted in Fig. \ref{fig:powerFig}. Note that the
dimensionless power spectrum is almost scale invariant between the scales
$1/l<k<1/w$ corresponding to the wake length and the wake width. When $k<1/l$
it decays as $k^2$, and when $k>1/w$ it decreases as $k^{-1}$. 

As an example, we consider $(G\mu)_6=0.2$ (around the
current upper bound on the string tension), $z_e=20$ and $N_H=10$ strings per Hubble
volume. From Fig.~\ref{fig:steps} we see that only wakes from a single Hubble time step
undergo shock heating. The relevant length scales of the wakes are
\begin{align}
  w=2.3 \times 10^{-4}{\rm Mpc}~, \qquad l=2.1{\rm~Mpc}~.
\end{align}
The number density of wakes from Eq.~\ref{n2D}, and the induced 21~cm brightness temperature from Eq.~\ref{eq:deltaTb} are
\begin{align}
  n_{2D}=0.12 {\rm~Mpc}^{-2}~, \qquad   \overline{\delta T_b} = -15{\rm~mK}~.
\end{align}
In this case, the dimensionless power spectrum becomes
\be
  \Delta(k) \, \simeq \, 8000 {\rm \mu K}^2 \qquad \mbox{for } 1/l < k < 1/w ~.
\ee
Comparing this result with the square of the local brightness temperature
from (\ref{eq:deltaTb}) we see that the power spectrum is suppressed by a
significant factor. This is a reflection of the highly non-Gaussian
distribution of wakes in space, and the resulting highly non-Gaussian
distribution of the string-induced wedges in the 21cm brightness maps.
The fraction of the 21cm sky at a fixed redshift which is covered by 
string-induced wedges is proportional to $w/l \sim G \mu$. Thus, whereas
the local brightness temperature inside a 21cm wedge is to a first approximation 
independent of the cosmic string tension, the two-dimensional power
spectrum decreases in amplitude as $G \mu$ decreases. The dependence
on $G \mu$ is linear as can be seen from (\ref{eq:Delta1}), since in the flat portion
of the spectrum the last term in the square brackets dominates.

Finally we also would like to comment which features of the power spectrum are expected
to exist for a realistic cosmic string network and which features are 
artifacts of the toy model. 

The decay rates of $\Delta(k)$ at small and large values of $k$ should
be robust because they do not depend on special assumptions made in the cosmic
string toy model. We  also trust the nearly scale invariant behavior between
$1/l$ and $1/w$. 

However, the oscillations seen in the spectrum near $k = 1/l$ are probably an
artifact of the toy model. This is because we have assumed all string wakes are
laid down at one particular redshift $z_m$. This is not accurate (up to an order
one factor) even in the case that only one Hubble step is needed. String wakes laid
down at different $z_m$ are expected to lead to cancellations because of the
different phases of oscillation. We also expect that the spectrum should have a 
slight blue tilt near $k = 1/l$. This is
because the larger scale wakes are laid down at later times and thus lead
to a lower brightness temperature. 

\subsection{Multiple Hubble steps}
\label{sec:mult-hubble-steps}

We now consider the case when there are several Hubble steps
which lead to wakes with $T_K < 2.5 T_g$. In this situation, the power spectrum 
becomes a superposition of the power spectra in different Hubble steps. For 
example, when we are considering measurements for emission redshift $z_e = 20$, 
then when $(G\mu)_6 > 0.22$, multiple Hubble steps are necessary. 

\begin{figure}[htpb]
\centering
\includegraphics[width=0.45\textwidth]{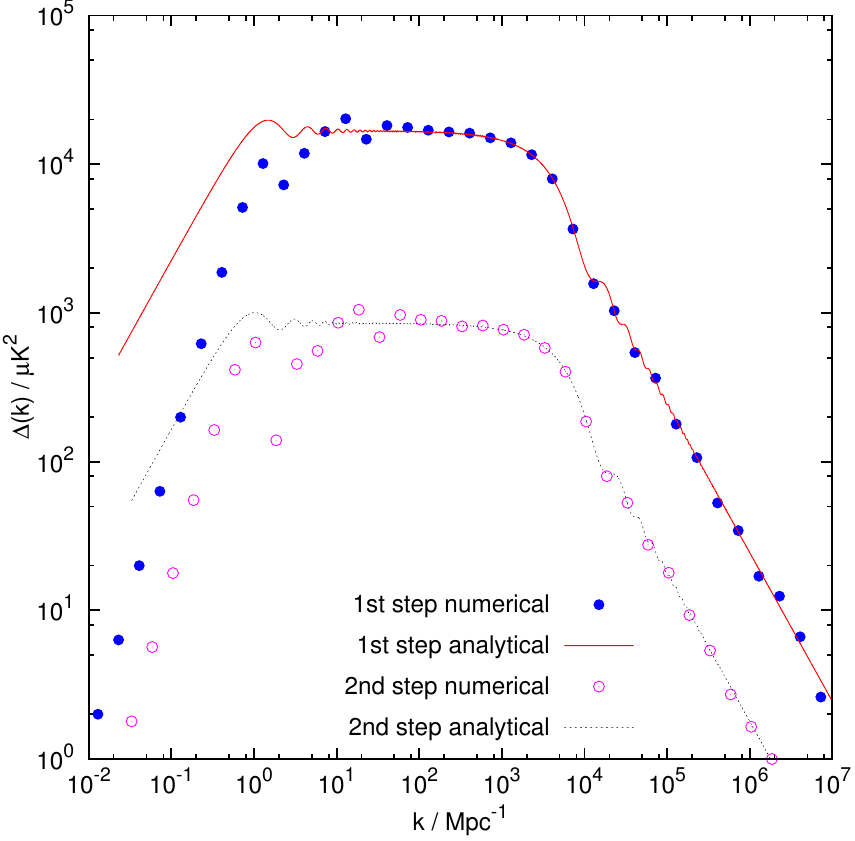}
\hspace{0.05\textwidth}
\includegraphics[width=0.45\textwidth]{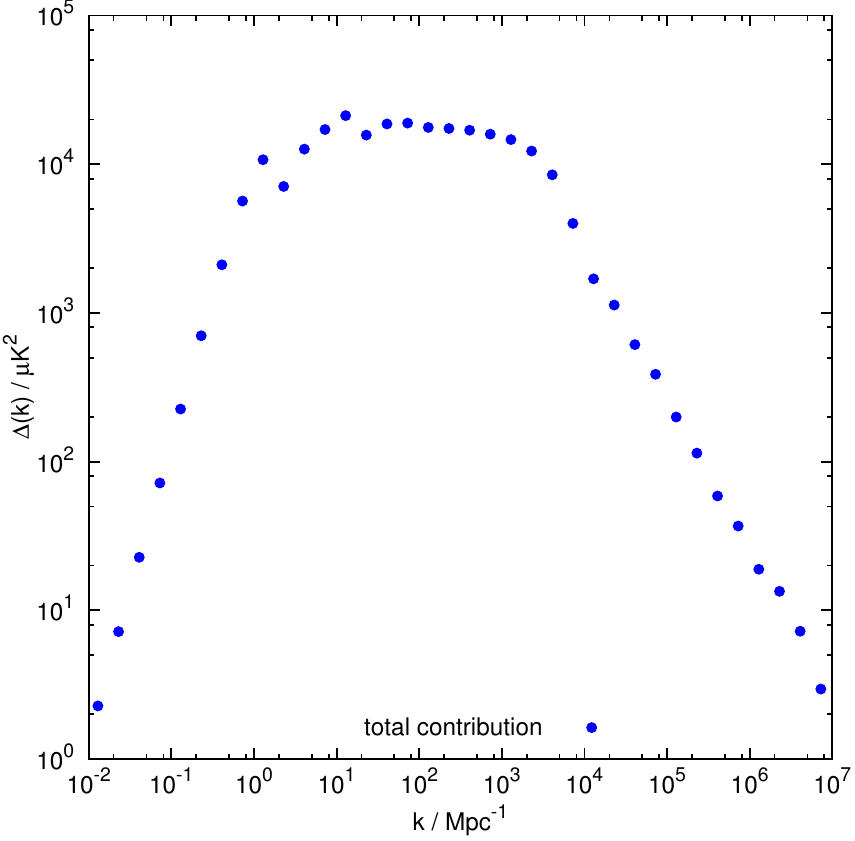}
\caption{\label{fig:stepFig} The dimensionless power spectrum from string wakes
at individual Hubble steps (left), and the sum over the steps (right). The 
parameters are $z_e=20$, $z_m^{\rm max}=3000$,
$(G\mu)_6=0.3$, $v_s=1/2$, $\cos^2\theta_s=1/2$.
and $N_H=10$. }
\end{figure}

We assume no cross correlation between string wakes in different Hubble
steps. Then, the contribution from different Hubble steps can be added together
directly. In the left and right panels of Fig. \ref{fig:stepFig}, we plot the
dimensionless power spectrum from wakes at each Hubble step and the sum,
respectively. Note that the first Hubble step has the largest brightness
temperature. Thus the first step dominates the total power spectrum.

\section{Conclusion}
\label{sec:conclusion}

We have computed the angular power spectrum of 21~cm emission (at a fixed
redshift) from a scaling distribution of string wakes. Wakes from strings 
present near the time of equal matter and radiation, dominate 
the spectrum. The dimensionless power spectrum scales as $k^2$ on long
distance scales and as $k^{-1}$ on short distances. In the above, ``long''
is relative to the length of the string wake which corresponds to an
angular scale of about $0.1^{\circ}$ and which is independent of the string
tension. ``Short'' is relative to the width of the wake which depends
on the string tension (its order of magnitude is $G \mu$ times the length
of the wake. On intermediate scales, the spectrum is roughly scale-invariant.

The above features of the 21~cm power spectrum from string wakes are robust
against details of the string wake scaling solution. The overall amplitude,
however, depends quite sensitively on the details. As an example, let us
consider the redshift $z_e = 20$, larger than the redshift
of reionization so that we do not need to worry about ``noise'' from
reionization processes. Then, for $(G \mu)_6 = 0.2$ (the
current upper bound on the string tension, and assuming 10 string segments
per Hubble volume, the amplitude of the dimensionless power spectrum
over the flat part of the spectrum is of the order 
$\Delta(k)^{1/2} \simeq 90~{\rm \mu K}$. 
This amplitude is significantly
smaller than the local brightness temperature in a string-induced 21cm wedge
in position space. Whereas the local brightness temperature in
position space is to a first approximation independent of the string
tension, the two-dimensional power spectrum decreases linearly
with $G \mu$ as the string tension decreases.
This difference between the local and the root mean
square power of the signal is a reflection of the highly non-Gaussian
distribution of cosmic string wakes, and it reinforces the lesson learned
from studies of cosmic string imprints on CMB temperature maps -
namely that it is easier to detect the signatures of cosmic strings when
analyzing the data in position space.

It will be of great interest to study in detail the detectability of this signal
with upcoming and planned experiments. To put the predicted amplitude
into some context, let us recall that the current strongest upper bound
on a cosmological 21~cm signal is $70~{\rm mK}$ \cite{GMRT} on angular
scales of about $3.3^{\circ}$. The redshift range probed in this experiment
(GMRT) lies between $8.1$ and $9.2$. One foreground for our signal is the
power spectrum induced by mini-halos, nonlinearities which result from
the Gaussian primordial fluctuations. An analytical estimate of the
effect was given in \cite{Pen1}, with the predicted spectrum $\Delta(k)^{1/2}$
at redshift $z = 15$ peaking at $k \simeq 10^3 h {\rm Mpc}^{-1}$ at an amplitude of
about $100~{\rm mK}$ and decaying linearly with $k$ for smaller values of $k$.
A detailed numerical study of the expected signal of reionization effects at a redshift
of $z = 11$ was performed in \cite{Pen2} yielding a spectrum which peaks
at $10 {\rm mK}$ at a value of the angular quantum number $l$ of $l \sim 10^4$
and decaying roughly linearly with $l$ for smaller values of $l$, and turning flat
for larger values. Turning to low reshift surveys, we recall that 21~cm experiments
are planned to be able to pick out a $20 {\rm \mu K}$ signal from baryon
acoustic oscillations from a noise of $300 {\rm mK}$ from large-scale
stucture on a length scale of about $8 {\rm Mpc}$ \cite{Peterson}.
Thus, it appears to us that the predicted spectrum from a scaling distribution
of cosmic string wakes is large and has a good chance of being
visible in the planned high redshift surveys.

Our work is based on a simple toy model for the distribution of
cosmic strings which captures the main properties of string
networks. Cosmic string loops and wiggles on long strings are,
however, not contained in the present toy model. It would be
of interest to extend our work to capture these aspects of
real string networks.
 
The restriction to fixed emission redshift cuts down a lot of the information
which is contained in 21~cm redshift maps. Hence, it will be of great
interest to extend our analysis and calculate the three dimensional
redshift power spectrum of 21~cm emission from cosmic strings. Work on
this topic is ongoing.

\section*{Acknowledgment}

This research is supported in part by an NSERC Discovery Grant, by
funds from the CRC program, and by a Killam Research Fellowship to R.B..
Y.W. acknowledges support from grants from
McGill University, Fonds Quebecois de la Recherche sur la Nature et
les Technologies (FQRNT), the Institute of Particle Physics (Canada) 
and the Foundational Questions Institute. O.H. is supported in part by
the FQRNT Programme de recherche pour les enseignants de coll\`ege. 
We are grateful for useful discussions with Rebecca Danos, Matt Dobbs, 
Gil Holder and Ue-li Pen. We thank Rebecca Danos for producing Figure 1 for us.


\end{document}